\documentclass[final,1p,times]{elsarticle}

\usepackage{amssymb}

\usepackage{amsthm}

\usepackage{amsmath}

\usepackage{xcolor}
\usepackage{hyperref}

\newcommand{\R}{\mathcal R}

\newcommand{\tond}[1]{{\left(#1\right)}}
\newcommand{\quadr}[1]{{\left[#1\right]}}
\newcommand{\inter}[1]{{\left\langle#1\right\rangle}}
\newcommand{\graff}[1]{\{#1\}}

\newcommand{\Poi}[2]{\{#1,#2\}}
\newcommand{\derp}[2]{{\frac{\partial #1}{\partial #2}}}

\newcommand{\norm}[1]{\left\|#1\right\|}
\newcommand{\normsup}[2]{\Big|#1\Big|_{#2}}

\newcommand{\Oscr}{\mathcal{O}}

\newcommand{\Rscr}{\mathcal{R}}

\def\CC{\mathbb{C}}
\def\RR{\mathbb{R}}
\def\ZZ{\mathbb{Z}}

\def\Id{\mathbb I}
\def\eps{\epsilon}

\def\im{i}

\def\Bscr{\mathcal{B}}
\def\Gscr{\mathcal{G}}

\def\Jscr{\mathcal{J}}

\def\Oscr{\mathcal{O}}

\def\Id{\mathbb I}

\def\epsilon{\varepsilon}
\def\eps{\varepsilon}

\def\derp#1#2{\frac{\partial#1}{\partial#2}}

\newtheorem{theorem}{Theorem}[section]
\newtheorem{theorem*}{Theorem}
\newtheorem{lemma}{Lemma}[section]

\newtheorem{proposition}{Proposition}[section]

\newtheorem{remark}{Remark}[section]

\begin{document}

\begin{frontmatter}



\title{Darboux's Theorem, Lie series and the standardization of the
  Salerno and Ablowitz-Ladik models\tnotemark[{label1}]\tnotemark[{label2}]}
\tnotetext[label1]{\color{red}The present version is the preprint one; for the accepted version on Physica D, please refere to the official page \url{https://doi.org/10.1016/j.physd.2024.134183}}
\tnotetext[label2]{©$\ $2024. This manuscript version is made available under the CC-BY-NC-ND 4.0 license; for detail please see \url{https://creativecommons.org/licenses/by-nc-nd/4.0/}}

\author[2]{Marco Calabrese}
\ead{Mcalabrese@umass.edu}
\author[1,3]{Simone Paleari}
\ead{simone.paleari@unimi.it}
\author[1,3]{Tiziano Penati\corref{cor1}}
\ead{tiziano.penati@unimi.it}
\cortext[cor1]{Corresponding author}

\affiliation[1]{organization={Department of Mathematics ``F.Enriques'', University of Milan},
            addressline={via Saldini 50}, 
            city={Milan},
            postcode={20133}, 
            country={Italy}}
\affiliation[2]{organization={Department of Mathematics and Statistics, University of Massachussets,},
            city={Amherst},
            postcode={01003-9305}, 
            state={MA},
            country={USA}}
\affiliation[3]{organization={GNFM (Gruppo Nazionale di Fisica Matematica) -- Indam
  (Istituto Nazionale di Alta Matematica ``F. Severi'')},
            city={Roma},
            country={Italy}}

\begin{abstract}
In the framework of nonlinear Hamiltonian lattices, we revisit the
proof of Moser-Darboux's Theorem, in order to present a general scheme
for its constructive applicability to Hamiltonian models with
non-standard symplectic structures. We take as a guiding example the
Salerno and Ablowitz-Ladik (AL) models: we justify the form of a
well-known change of coordinates which is adapted to the Gauge
symmetry, by showing that it comes out in a natural way within the
general strategy outlined in the proof.  Moreover, the full or
truncated Lie-series technique in the extended phase-space is used to
transform the Salerno model, at leading orders in the Darboux
coordinates: thus the dNLS Hamiltonian turns out to be a normal form
of the Salerno and AL models; as a byproduct we also get estimates of
the dynamics of these models by means of dNLS one.
We also stress that, once it is cast into the perturbative approach,
the method allows to deal with the cases where the explicit
trasformation is not known, or even worse it is not writable in terms
of elementary functions.

\end{abstract}







\end{frontmatter}

\section{Introduction}
\label{s:intro}

On a symplectic manifold $(M,\omega)$ of dimension $2n$, Darboux's
Theorem of symplectic geometry \cite{Dar1882} ensures the local
existence of a set of coordinates, say $(q_j,p_j),\, j=1,\ldots,n$,
such that at any point $P\in M$ the symplectic 2-form $\omega$ reads
$\omega_P = \sum_j dq_j\wedge dp_j$.  The existence of such a standard
set of coordinates is useful for the Hamiltonian formalism, and for
Hamiltonian perturbation theory in particular (see \cite{Gio22}),
since Hamiltonian equations, Hamiltonian vector fields $X_H$,
symmetries and canonical transformations can be expressed in terms of
Poisson brackets $\Poi{\cdot}{\cdot}$ via the (constant) symplectic
matrix $J$ as follows
\begin{equation}
  \label{e.J}
  \dot z_j=\left(X_H\right(z))_j=\Poi{z_j}{H}
  \qquad\quad
  \Poi{f}{g}=(\nabla f)^\top J\nabla g\ ,
  \qquad
  J=\begin{pmatrix}
  0 & \Id\\
  -\Id & 0
  \end{pmatrix}\ ,
\end{equation}
where $\Id$ represents the $n$-dimensional identity matrix and the
gradient $\nabla$ is intended with respect to the set of real
coordinates $z=(q_j,p_j)$.

A classical example, in the field of Hamiltonian Lattices (see
\cite{PalP19} for a recent review on the topic), is the
discrete\footnote{It is indeed the most common discretization of the
  continuous Nonlinear Schroedinger Equation (NLS) model.}  Nonlinear
Schro\"dinger model (dNLS)
\begin{flalign}
    \label{e.dNLS.eq}
    \textrm{\bf [dNLS]} &&
    \im \dot \psi_j = \eps(\psi_{j+1} +\psi_{j-1}) + \gamma|\psi_j|^2\psi_j\ ,
    &&
\end{flalign}
where $\psi_j\in\CC$ and the lattice-index $j$ may run in a finite
$\Jscr=\graff{1,\ldots,N}$ (with periodic or fixed boundary
conditions) or infinite $\Jscr=\ZZ$ sets (with $\ell^2$-decay of the
sequence $\graff{\psi_j}_{j\in\Jscr}$). It turns out ot be an
Hamiltonian system with Hamiltonian given by
\begin{flalign}
    \label{e.dNLS}
    \textrm{\bf [H}_\textrm{\bf dNLS}\textrm{\bf ]} &&
    H(\psi) = \sum_{j\in\Jscr} \quadr{\eps(\psi_{j+1}\bar\psi_j + \bar\psi_{j+1}\psi_j)
      + \frac\gamma2|\psi_j|^4}\ ,
    &&
\end{flalign}
with the {\em standard Poisson structure} $\Poi{\psi_j}{H} =
-\im\derp{H}{\overline\psi_j}$ in complex variables.

However, it might happen that the dynamics of a physical model is
described by a vector field $X_H(z)$, with physical coordinates $z$,
which is derived by the Hamiltonian $H(z)$ through a non-standard
representation of the Poisson brackets $\Poi{\cdot}{\cdot}$; in this
case the 2-form $\omega$ is locally represented by a different
(typically non constant) matrix $\Omega(z)$, such that
$\omega_P(X,Y)=X^\top \Omega(z)Y$, being $\Omega(z)$ antisymmetric and
non-degenerate. This is the case of the Ablowitz-Ladik (AL in the
following) system
\begin{flalign}
  \label{e.AL.eq}
  \textrm{\bf [AL]} &&
  \im  \dot \psi_j = \eps(1+\mu|\psi_j|^2)(\psi_{j+1} + \psi_{j-1})\ ,
  &&
\end{flalign}
which is a celebrated integrable discretization of the NLS, and of the
Salerno models
\begin{flalign}
  \label{e.Sal.eq}
  \textrm{\bf [Salerno]} &&
  \im \dot \psi_j = \eps(1+\mu|\psi_j|^2)(\psi_{j+1} +\psi_{j-1}) +
  \gamma|\psi_j|^2\psi_j\ ;
  &&
\end{flalign}
the two common parameters $\eps$ and $\mu$ can be taken as positive,
while $\gamma$ of any sign. 
It is well known that both models are Hamiltonian and share the same
{\em nonstandard symplectic structure} given by the Poisson brackets
\begin{equation}
  \label{e.Poi.nonst}
  \dot\psi_j = (X_H)_j(\psi) = \Poi{\psi_j}{H}\qquad\qquad
  \Poi{\psi_j}{H} = -\im(1+\mu|\psi_j|^2)\derp{H}{\overline\psi_j}\ ,
\end{equation}
while the Hamilton function reads
\begin{equation}
  \label{e.ALSal.Ham}
  H(\psi) = \sum_{j\in\Jscr} \quadr{-\frac\gamma{\mu^2}\ln(1+\mu|\psi_j|^2) + \eps
  (\psi_{j+1}\bar\psi_j + \bar\psi_{j+1}\psi_j) + \frac\gamma\mu|\psi_j|^2}\ ;
\end{equation}
for $\gamma=0$ we recover the AL model \eqref{e.AL.eq} while
$\gamma\neq 0$ gives the Salerno model \eqref{e.Sal.eq}.

The two models also share a second conserved quantity
\begin{equation}
  \label{e.P}
P=\frac1\mu\sum_{j\in\Jscr}\ln(1+\mu|\psi_j|^2)\qquad\qquad \Poi{H}{P}=0\ ,
\end{equation}
which is related to the Gauge symmetry $e^{\im \theta}$ of the
equation; indeed, the flow of the Hamiltonian vector field $(X_{P})_j
= \Poi{\psi_j}{P} = -\im\psi_j$ is exactly given by the action of
$e^{\im \theta}$.

One of the reasons why it is preferable to standardize the symplectic
structure \eqref{e.Poi.nonst} --- which by the way coincides with the
standard one when $\mu=0$ --- is the possibility of linking
homotopically the AL to the dNLS at the Hamiltonian level, by setting
$\gamma=1-\mu$, with $\mu\in [0,1]$; indeed the Salerno model
interpolates between the AL model (for $\gamma=0$) and the dNLS (for
$\gamma\neq 0$ and $\mu\to 0$), and a common symplectic structure
might be preferable in order to explore the transition
(AL-Salerno-dNLS) along the three models, especially in terms of
dynamical features (see \cite{HenKC22,HenKC22b} for a comparison
between AL and dNLS in terms of vector fields and persistence of
localized structures, see instead \cite{Mit_etal_23} for a comparison
between AL and Salerno in terms of additional conserved
quantities). On the other hand, standardization of \eqref{e.Poi.nonst}
might be helpful to implement geometric numerical schemes which are
more suitable for the integrability of AL (see for example
\cite{TangLCS07}).

Motivated by the above mentioned issues concerning those Hamiltonian
non linear lattices, the present manuscript aims at providing a new
insight into the classical problem of applying Darboux's Theorem to
specific models, when a given Hamiltonian system has to be explicitly
transformed into Darboux's coordinates.

As a first result (see Theorem \ref{t.main.1}) we show that one of
most used Darboux's transformation for AL and Salerno models (see
again \cite{TangLCS07} and \cite{HeiLW02}), i.e.
\begin{equation}
  \label{e.D_trsf.compl}
  \psi_j = \Psi_j \sigma\left(\frac\mu2|\Psi_j|^2\right)
  \qquad\qquad\textrm{with}\quad
  \sigma(s) = \sqrt{\frac{\exp(s)-1}{s}}\ ,
\end{equation}
can be derived directly from Moser's scheme of the proof (originally
in \cite{Mos65}, here taken from \cite{Lee12}). We implement such a
transformation with the equivalent procedure of the Lie-series (see
the classical works \cite{Dep69, Hor66, Grob73}): this method relies
on the idea that, since the change of coordinate suggested by the
proof is the flow of a vector field $V$ at a given time $t=1$, the
transformed Hamiltonian systems can be obtained as a (totally
convergent) series of iterated Lie derivatives $L_VH$ along $V$. In
the scheme of Moser, such a vector field is time dependent, hence the
Lie-series representation of \eqref{e.D_trsf.compl} is performed in
the extended $\RR\times\RR^{2n}$ phase-space
\begin{equation*}
H(\Psi\sigma(\Psi)) = \exp\tond{-L_{\tilde V}H}(\Psi) = \sum_{l\geq 0}
(-1)^l L_{\tilde V}^lH(\Psi) \ ,
\end{equation*}
where $L_{\tilde V}H$ is the Lie-derivative of $H$ along the extended
vector field $\tilde V=(1,V)$.

As a second result we quantitatively compare the models under
investigation, both at the level of the Hamiltonians, in the spirit
of normal forms, and at the level of solutions.  Indeed, once put in
the transformed coordinates $\Psi$, the AL and Salerno model can be
compared to the dNLS dynamics by exploiting the Hamiltonian formalism,
since they all share the same symplectic structure, at least in a
small neighborhood of the origin. We thus provide estimates of the
closeness between the AL and the dNLS model, as well as between the
Salerno model \eqref{e.ALSal.Ham_r}, or a suitable cubic-quintic
generalization (see Theorem \ref{t.main.2} for a more precise
formulation), and the dNLS: in all the cases small norm initial data
have to be considered, since the Darboux transformation is only
local. Similar results have been already obtained in \cite{HenKC22}:
however, at variance with these authors, we here make estimates at the
level of the Hamiltonian formalism and with the use of Lie-series,
rather than directly controlling the difference between the two vector
fields.

A further result pertains the additional conserved quantity that all
the models we are considering admit.  With the above transformation
\eqref{e.D_trsf.compl}, the AL, Salerno and dNLS models now share the
same conserved quantity, since the quantity $P$, defined in
\eqref{e.P}, becomes the $\ell^2$ norm (which is the additional
conserved quantity of the dNLS model \eqref{e.dNLS}). In terms of
Lie-series, the integral given by the $\ell^2$ norm has to coincide
(modulo a prefactor) with $\exp(-L_{\tilde V}P)(\Psi)=\frac12
\norm{\Psi}^2$.
It is possible to prove (see Proposition \ref{t.main.3}) --- and
easily anticipate by means of a numerical evidence (using Mathematica)
--- a geometric convergence of the truncated Lie-series
$\exp_K(-L_{\tilde V}P)(\Psi)=\sum_{l=0}^K (-1)^l L_{\tilde
  V}^lH(\Psi)$ to the $\ell^2$ norm, for increasing values of
$K$. Indirectly, this also numerically confirms the correct
relationship between \eqref{e.D_trsf.compl} and the (time dependent)
vector field $V$ explicitly computed going through Moser's scheme.

Clearly, in these specific applications, in particular for the second
result, we might have avoided the use of Lie-series, being in fact
equipped with the explicit form of $\Psi$. However, in the spirit of
perturbation theory, we stress that if one is interested only in some
leading order approximation (normal form) of the transformed Hamilton
function $H(\Psi\sigma(\Psi))$, it is enough to simply construct a
proper polynomial approximation $\tilde V^{(L)}$ of $\tilde V$, which
can be then used to transform $H$ through the (truncated) Lie-series
$\exp_K(L_{\tilde V^{(L)}}H)$; it is indeed possible, with some
standard analytical estimates of Lie-series, to derive a priori bounds
of the error $|\exp(L_{\tilde V }H)-\exp_K(L_{\tilde V^{(L)}}H)|$. And
in fact such a procedure can be exploited even in cases where the
explicit form of the transformation is not known or does not exist in
terms of elementary functions (see \cite{JungMW17} for a related
approach).

\bigskip

The scheme of the manuscript is the following.  Section 2 will be
devoted to present the results of this paper: we show the emergence of
the aforementioned transformation from the abstract geometric scheme
of the proof and provide normal form statements in the framework of
the Lie-series. Section 3 includes the proofs of the results. Some
additional comments and future perspectives are included in the
Conclusions, Section 4. In Appendix A we recall some analytical
results on Lie-series in the non-autonomous case. In Appendix B we
review Moser's proof and we extract the constructive scheme.


\section{Results}
\label{s:res}

In order to find a set of Darboux coordinates, according to the scheme
of the proof here reviewed, we prefer to work with a set of canonical
and cartesian variables $\graff{q_j,p_j}\in\RR^{2N}$ defined by
\begin{displaymath}
  \psi_j=\frac1{\sqrt2}(q_j+\im p_j)\qquad\qquad
  \overline\psi_j=\frac1{\sqrt2}(q_j-\im p_j)\ ;
\end{displaymath}
from now on we assume $\Jscr$ finite with cardinality $N$, but the
results here presented work also in the infinite lattice. We start
rewriting the Poisson brackets of two smooth functions $F,\,G$ as
\begin{equation*}
\Poi{F}{G} = \sum_j \quadr{1+\nu(q_j^2+p_j^2)}\graff{\partial_{q_j}
  F\partial_{p_j} G - \partial_{p_j} F\partial_{q_j} G}\ ,
\end{equation*}
where $\nu=\frac12\mu$ and the Poisson brackets are related to the
2-form $\omega_0$
\begin{equation*}
\omega_0(q,p) = \sum_j\frac1{1+\nu(q_j^2+p_j^2)}dq_j\wedge dp_j\ .
\end{equation*}
We notice that at the origin $\omega_0$ coincides with the
standard\footnote{As will be briefly explained in Appendix B, it is
  always possible to set $\omega_0(P)=\omega_1$ at a given point $P$,
  with a linear change of coordinates.} symplectic form
$\omega_1$
\begin{equation}
\label{e.omega1}
  \omega_1 = \sum_j dq_j\wedge dp_j = \omega_0(0,0)\ .
\end{equation}

The AL and Salerno Hamiltonian can be then rewritten in cartesian
variables
\begin{equation}
  \label{e.ALSal.Ham_r}
  H = H_0 + H_1\qquad
  \begin{cases}
    H_0 = \phantom{\eps}\sum_{j\in\Jscr}
    \quadr{-\frac\gamma{4\nu^2}\ln[1+\nu
        (q_j^2+p_j^2)] + \frac\gamma{4\nu}(q_j^2+p_j^2)}
    \\
    H_1 = \eps\sum_{j\in\Jscr} (q_{j+1}q_j + p_{j+1}p_j)
  \end{cases}
  ,
\end{equation}
and the corresponding Hamilton equations are
\begin{equation*}
  \begin{cases}
    \dot q = \phantom{-}\eps(p_{j+1} + p_{j-1})\tond{1+\nu(q_j^2+p_j^2)} +
    \frac12 p_j\gamma(q_j^2+p_j^2)
    \\
    \dot p = -\eps(q_{j+1} + q_{j-1})\tond{1+\nu(q_j^2+p_j^2)} -
    \frac12 q_j\gamma(q_j^2+p_j^2)\ ;
  \end{cases}
\end{equation*}
in the same set of coordinates, the Hamiltonian of the dNLS
model \eqref{e.dNLS.eq} reads
\begin{equation}
  \label{e.dNLS.Ham_r}
  H = H_0 + H_1
  \qquad\qquad
  \begin{cases}
    H_0 = \frac\gamma{8}\sum_{j\in\Jscr} (q_j^2+p_j^2)^2
    \\
    H_1 = \eps\sum_{j\in\Jscr} (q_{j+1}q_j + p_{j+1}p_j)
  \end{cases}
  \ ,
\end{equation}
with Hamilton equations given by
\begin{equation*}
  \begin{cases}
    \dot q = \phantom{-}\eps(p_{j+1} + p_{j-1}) +
    \frac12 p_j\gamma(q_j^2+p_j^2)
    \\
    \dot p = -\eps(q_{j+1} + q_{j-1}) -
    \frac12 q_j\gamma(q_j^2+p_j^2)\ ;
  \end{cases}
\end{equation*}

Since the Darboux transformation acts in the same way on each
two-dimensional subspace with coordinates $(q_j,p_j)$, we can restrict
to the 2-dimensional manifold $\RR^2$ and consider $\omega_0$ as
\begin{equation*}
\omega_0 = \frac1{1+\nu(q^2+p^2)}dq\wedge dp\ ,
\end{equation*}
by omitting all the indexes; in this way, denoting again by $J$ the
restriction of \eqref{e.J} on $\RR^2$, one has
\begin{equation}
  \label{e.Omega}
\Omega(q,p) = \frac1{1+\nu(q^2+p^2)}J\ ,
\end{equation}
and the additional conserved quantity \eqref{e.P} takes the form (on
the 2-dimensional subspace)
\begin{equation}
  \label{e.P_r}
  P(q,p) = \frac1{2\nu}\ln[1+\nu (q^2+p^2)]\ .
\end{equation}
As anticipated in~\eqref{e.D_trsf.compl}, the nonlinear change of
coordinates $(q,p)=\varphi^{-1}(x,y)$ with $\varphi^{-1}$ given by
\begin{equation}
  \label{e.D_trsf}
  \begin{cases}
    q = x \sigma(\nu\norm{(x,y)}^2)\\
    p = y \sigma(\nu\norm{(x,y)}^2)\\
  \end{cases}
  \ ,
\end{equation}
is a Darboux transformation\footnote{Indeed it satisfies $J =
  (D_{x,y}\varphi)^\top \Omega(\varphi(x,y)) (D_{x,y}\varphi)$.}  and
it transforms $P$ into the $\ell^2$ norm (modulo a prefactor
$\frac12$)
\begin{equation}
  \label{e.P_Dar}
  P\circ\varphi^{-1}(x,y) = \frac12(x^2+y^2) .
\end{equation}

\medskip
\subsection{First result: Darboux's change of coordinates}

As a first main result we show the relationship between
\eqref{e.D_trsf} and Moser's scheme of the proof, where the last can
be summed up in the three main steps:
\begin{enumerate}

\item find a vector potential $a(q,p)=(a_1,a_2)$ for the closed 2-form
  $\omega_0-\omega_1$, namely  such that
  \begin{equation}
    \label{e.a}
    \partial_p a_1 - \partial_q a_2 = \frac{\nu(q^2+p^2)}{1+\nu(q^2+p^2)}\ ;
  \end{equation}
  
\item compute the vector field $V_t(q,p)$
  \begin{equation}
    \label{e.Vt}
    V_t(q,p) = \Omega_t^{-\top}(q,p) a(q,p)\ ,
  \end{equation}
  where, by taking $t\in[0,1]$, the matrix $\Omega_t(q,p)$
  interpolates between $\Omega(q,p)$ (non-standard symplectic
  structure given by \eqref{e.Omega}) and $J$ (the standard one)
  \begin{equation*}
    \Omega_t(q,p)=tJ +(1-t)\Omega(q,p)\qquad\qquad
    \Omega_1=J\qquad\qquad\Omega_0=\Omega\ .
  \end{equation*}

\item solve (if possible) the dynamical system
  \begin{equation*}
    (\dot q,\dot p) = V_t(q,p)\ ,
  \end{equation*}
  whose time-one-flow $\Phi^1(q,p)$ defines the Darboux transformation
  $(x,y) = \varphi(q,p)=\Phi^1(q,p)$.
\end{enumerate}

In the following, we make use of the Lie-series formalism to represent
$\Phi^1(q,p)$; in particular, being the vector field $V_t$ time
dependent, we introduce the extended vector field $\tilde V=(1,V_t)$
defined in the extended phase space $(\tau,q,p)\in\RR^3$ and the
corresponding Lie-derivative of a analytic function $f:\RR^3\to\RR$
\begin{equation}
\label{e.Lie.op}
    L_{\tilde V} f = \partial_\tau f + \inter{V_t(q,p),\nabla f(q,p)}\ .
\end{equation}
The extended field $\tilde V$ defines a flow
$\tilde\Phi^t:\RR^3\to\RR^3$ which can be expressed in term of the
Lie-series operator $\exp(t L_{\tilde V})$
\begin{equation}
\label{e.flow.ext}
    \tilde\Phi^t(\tau,q,p) = \exp(t L_{\tilde V}\Id)(\tau,q,p) =
    \sum_{k\geq 0}\frac{t^k}{k!} (L_{\tilde V}^k \Id)(\tau,q,p)\ ,
\end{equation}
where $\Id:\RR^3\to\RR^3$ is the identity map and $L_{\tilde V}^k \Id$
has to be interpreted as $(L_{\tilde V}^k \Id)_j = L_{\tilde V}^k
\Id_j$, being $\Id_j$ the identity on the $j_{th}$ component.  It
turns out that
\begin{equation*}
    \tilde\Phi^t(0,q,p)= (t,\Phi^t(q,p))\qquad\Longrightarrow\qquad 
    \tilde\Phi^1(0,q,p) = (1,x,y)=(1,\varphi(q,p))\ ,
\end{equation*}
and consequently
\begin{equation*}
    \tilde\Phi^{-1}(1,x,y) = \sum_{k\geq 0}\frac{(-1)^k}{k!}
    (L_{\tilde V}^k \Id)(1,x,y)= (0,q,p) = (0,\varphi^{-1}(x,y))\ .
\end{equation*}

\medskip

\begin{theorem}
  \label{t.main.1}
  Let $V_t(q,p)$ be the time dependent vector filed given by
  \begin{equation}
  \label{e.Vt_AL}
     V_t(q,p) =\chi (t,\nu\norm{(q,p)}^2)
    \begin{pmatrix}
      q \\
      p 
    \end{pmatrix}
      \end{equation}
  with 
  \begin{equation}
    \label{e.chi}
    \chi(t,\nu\norm{(q,p)}^2) =
    \frac{1+\nu\norm{(q,p)}^2}{2(1+t\nu\norm{(q,p)}^2)}
    \quadr{\frac{\ln(1+\nu\norm{(q,p)}^2)}{\nu\norm{(q,p)}^2}-1}\ ,
  \end{equation}
  and consider its extension $\tilde
  V(\tau,q,p)=(1,V_t(q,p)):\RR^3\to\RR^3$.  Then, the time-one-flow
  $\tilde\Phi^1(0,q,p)=(1,\varphi(q,p))$ generated by $\tilde V$ is
  well defined in a sufficiently small ball
  $B_\rho(0)=\{\norm{(q,p)}<\rho\}$ of the origin.  Moreover, the
  Darboux change of coordinates \eqref{e.D_trsf} corresponds to the
  inverse transformation $(q,p)=\varphi^{-1}(x,y)$, where $\varphi$ is
  a near-to-the-identity and analytic map with the following
  asymptotic expansion
\begin{equation}
\label{e.phi.exp}
\varphi(q,p) \sim \quadr{1 -\frac14 \nu\norm{(q,p)}^2}\begin{pmatrix}
      q \\
      p 
    \end{pmatrix}\ .
\end{equation}
\end{theorem}

\begin{remark}
  In agreement with \eqref{e.phi.exp}, we observe that
  \eqref{e.D_trsf} has the asymptotic expansion $\varphi^{-1}(x,y)
  \sim (x,y) +\frac14\nu\norm{(x,y)}^2 (x,y)$.
\end{remark}

Some other remarks on Theorem \eqref{t.main.1} are in order:
\begin{enumerate}

\item the vector field $V_t$ is equivariant under the same Gauge
  symmetry of the models \eqref{e.ALSal.Ham_r}. This has been made
  more evident by a proper choice of the vector $a(q,p)$, solution of
  \eqref{e.a}. Indeed the vector potential $a(q,p)$ in \eqref{e.a} is
  defined modulo a gradient $\nabla f$ of a scalar function $f$;

\item in general, we cannot expect to write the flow of the dynamical
  system given by $V_t$; but we can prove that in a small
  neighbourhood of the origin the flow $\Phi^t(q,p)$ is radial and
  close to the identity (and contracting).  Hence, we can impose a
  precise structure to the unknown transformation $\Phi^1(q,p)$ (and
  to its inverse), which leads to \eqref{e.D_trsf} as its unique
  analytic solution;

\item the vector field $V_t$ is asymptotically cubic for $(q,p)$ in a
  sufficiently small neighborhood of the origin. This can be
  understood by observing that $V_t$ is constructed from the nonlinear
  deformation of $\omega_0$ with respect to $\omega_1$. Indeed the
  vector $a(q,p)$ is obtained integrating the quadratic deformation
  $\frac{\nu(q^2+p^2)}{1+\nu(q^2+p^2)}$ in \eqref{e.a}. As a
  consequence, the flow $\Phi^t(q,p)$ is a nonlinear deformation of
  the identity map in $B_\rho(0)$.

\end{enumerate}

\medskip
\subsection{Second result: dNLS-like normal forms}

The approach of Lie-series allows to transform any Hamiltonian system,
once given the generator vector field $V_t$. Indeed, the explicit
expression of the flow $\tilde\Phi^1$ is not necessary and the
Lie-series $\exp{(\pm L_{\tilde V})}$ operator suffices to write the
Hamilton equations in the transformed variables $(x,y)$. However, the
knowledge of $V_t$ depends on the possibility to provide the vector
potential $a(q,p)$ through an explicit integration: in the spirit of
perturbation theory, the leading order approximation of the
transformed Hamiltonian $\exp(-L_{\tilde V}H) = H\circ\varphi^{-1}$
can be obtained by a suitable truncation of $V_t$, hence by a
polynomial approximation of the vector potential $a(q,p)$, which is
always an accessible task. In order to formulate the next statement,
we introduce the following notation for the Taylor expansion of $H$ in
\eqref{e.ALSal.Ham_r} and of $V_t$ in \eqref{e.Vt_AL}:
\begin{equation*}
  H_0=\sum_{l\geq 2}H_{0,2l}
  \qquad\qquad
  H_1=H_{1,2}\ ,
  \qquad\qquad
  V_t=\sum_{l\geq 1} V_{t,2l+1}\ .
\end{equation*}
where the second index $s$ in $H_{j,s}$ represents the polynomial
degree of $H_{j,s}$ in the variables $(q,p)$, while the first index
$j$ is the degree with respect to the parameter $\eps$. Furthermore,
$\tilde V_{2l+1}$ will denote the extension of $V_{t,2l+1}$ and
$\Phi^t_H(x_0,y_0)$ denotes the Hamiltonian flow associated to $H$ at
time $t$, with initial datum $(x_0,y_0)\in\RR^{2n}$.  

\begin{theorem}
  \label{t.main.2}
  In a sufficiently small ball $B_\rho(0)\subset\RR^{2n}$ of the
  origin, the Hamiltonian \eqref{e.ALSal.Ham_r} can be transformed by
  the inverse Lie-series along $\tilde V$
  \begin{equation*}
    H\circ\varphi^{-1}(x,y) = \exp(-L_{\tilde V}H) =
    \sum_{k\geq 0}\frac{(-1)^k}{k!} (L^k_{\tilde V} H)(1,x,y)\ .
  \end{equation*}
  At leading order (in $\rho$ and $\eps$), the Salerno model
  \eqref{e.ALSal.Ham_r} can be approximated by the dNLS Hamiltonian
  \eqref{e.dNLS.Ham_r}
  \begin{equation*}
    \begin{aligned}
      \exp(-L_{\tilde V}H) &= Z^{(0)} + \Rscr^{(0)}\\ 
      Z^{(0)}&=H_{0,4} + H_{1,2} = \sum_{j\in\Jscr}\quadr{
        \frac\gamma8\tond{x_j^2+y_j^2}^2 +
        \eps\tond{x_{j+1}x_j+y_{j+1}y_j}}\ ,    
    \end{aligned}
  \end{equation*}
  where the remainder $\Rscr^{(0)}$ satisfies 
  \begin{equation*}
    \sup_{(x,y)\in B_\rho}|{\Rscr^{(0)}(x,y)}|\leq C_0\nu\rho^4({\rho}^2+\eps)\ ,
    \qquad C_0>0\ ,
  \end{equation*}
  and for times $|t|\leq (\rho^2+\eps)^{-1}$ one has
  \begin{equation}
    \label{e.Z0.cfr}
    \norm{\Phi^t_H(x_0,y_0)-\Phi^t_{Z^{(0)}}(x'_0,y'_0)}\leq
    c_0\tond{\norm{(x_0,y_0)-(x'_0,y'_0)}+\rho^3}\qquad\qquad  c_0>0\ .
  \end{equation}

  At next order, the Salerno model \eqref{e.ALSal.Ham_r} admits the
  following cubic-quintic normal form
  \begin{equation}
    \label{e.Z1}
    \begin{aligned}
      \exp(-L_{\tilde V}H) &= Z^{(1)} + \Rscr^{(1)}\\
      Z^{(1)} &= H_{0,4} + H_{1,2} - L_{\tilde V_3}H_{0,4} - L_{\tilde V_3}H_{1,2}= \\
      &= \sum_{j\in\Jscr}\quadr{
        \frac\gamma8\tond{x_j^2+y_j^2}^2 +
        \eps\tond{x_{j+1}x_j+y_{j+1}y_j}} + \\
      &+\sum_{j\in\Jscr}\quadr{\frac\gamma{24}\nu
        \tond{x_j^2+y_j^2}^3 + \frac14\eps\nu\tond{x_j^2+y_j^2}
        \tond{(x_{j+1}+x_{j-1})x_j+(y_{j+1}+y_{j-1})y_j}}\ ,  
    \end{aligned}  
  \end{equation}  
  where the remainder $\Rscr^{(1)}$ satisfies 
  \begin{equation*}
    \sup_{(x,y)\in B_\rho}|{\Rscr^{(1)}(x,y)}|
    \leq C_1\nu^2\rho^6({\rho}^2+\eps)\ ,\qquad C_1>0\ ,
  \end{equation*}
  and for times $|t|\leq (\rho^2+\eps)^{-1}$ one has
  \begin{equation}
    \label{e.Z1.cfr}
    \norm{\Phi^t_H(x_0,y_0)-\Phi^t_{Z^{(1)}}(x'_0,y'_0)}\leq
    c_1\tond{\norm{(x_0,y_0)-(x'_0,y'_0)}+\rho^5}
    \qquad\qquad  c_1>0\ .
  \end{equation}

  The dNLS Hamiltonian $Z^{(0)}$ is a normal form also for the AL model, with
  a remainder $\Rscr^{(0)}$ satisfying
  \begin{equation*}
    \sup_{(x,y)\in B_\rho}|{\Rscr^{(0)}(x,y)}|\leq C'_0\rho^4(1+\eps)\ ,
    \qquad C'_0>0\ ;
  \end{equation*}
  for times $|t|\leq \eps^{-1}$ one has
  \begin{equation}
    \label{e.Z0.AL.cfr}
    \norm{\Phi^t_H(x_0,y_0)-\Phi^t_{Z^{(0)}}(x'_0,y'_0)}\leq
    c'_0\tond{\norm{(x_0,y_0)-(x'_0,y'_0)}+\frac{\rho^3}{\eps}}
    \qquad\qquad  c'_0>0\ .
  \end{equation}
  
\end{theorem}

\medskip

Some remarks on Theorem \eqref{t.main.2} are also in order:
\begin{enumerate}

\item the dNLS approximation of the AL dynamics given by formula
  \eqref{e.Z0.AL.cfr}, for small (in norm) initial data, is in
  agreement with the first (and main) statement in \cite{HenKC22};
  indeed, for $\eps=\Oscr(1)$, our estimate \eqref{e.Z0.AL.cfr} claims
  that two initial data which are $\rho^3$-close and of order $\rho$,
  stay $\rho^3$-close for times of order $|t|\leq \Oscr(1)$. In this
  case, our proof can be adapted in order to consider a fixed time
  scale $|t|\leq T$ with arbitrary $T>0$; however, as in
  \cite{HenKC22}, the price to pay would be a constant $c'_0(T)$ which
  is increasing with $T$. At variance with \cite{HenKC22}, we do not
  need to impose any condition on $P(0)$, since the AL dynamics is
  compared to the dNLS one only after the Darboux transformation has
  been performed: hence the two models share the same conserved
  quantity, namely the norm. In fact, the smallness condition on
  $P(0)$ in \cite{HenKC22} is asked in order to ensure that the AL
  flow keeps its norm bounded for all times: this is obtained for free
  with our approach;

\item estimates \eqref{e.Z0.cfr} and \eqref{e.Z1.cfr} provide two
  different approximations of the Salerno model ($\gamma\neq 0$) on
  the same time scale $\Oscr((\rho^2+\eps)^{-1})$. The time scale
  suggests to link $\eps$ to $\rho$ in the regime $\eps\sim\rho^2\ll
  1$, so that the two statements are formulated only in terms of
  energy (or amplitude). This regime is the typical one for which the
  dNLS is the normal form of the Klein-Gordon chain (see for example
  \cite{PalP14,PelPP16});
  
\item it is clear by applying the transformation \eqref{e.D_trsf} to
  the model \eqref{e.ALSal.Ham_r} (by truncating the Taylor expansion
  at the identity) that the dNLS \eqref{e.dNLS} is the first
  approximation of \eqref{e.ALSal.Ham_r}, when $\gamma\neq
  0$. However, we here want to derive different levels of
  approximation of the model by exploiting the Lie-series method and
  the expansion of the vector field $V_t$, without any need of knowing
  the exact shape either of $\varphi$ or of $V_t$.  Furthermore, the
  correctness of the expansion of $H\circ \varphi^{-1}$ can be
  directly verified thanks to the explicit knowledge of $\varphi$ in
  \eqref{e.D_trsf};

\item as it is usual in the non autonomous case, we move to the
  extended phase space, so to apply the standard Lie-series operator
  \eqref{e.flow.ext} to transform the Hamiltonian in the new Darboux
  coordinates. Since by construction we already know the symplectic
  form in the new coordinates, it is enough to transform the
  Hamiltonian in order to get the Hamilton equations in the new set of
  variables;

\item as already stressed at the beginning of this subsection, we
  remark that, if we are interested in a leading order expansion of
  the transformed Hamiltonian $\exp(L_{\tilde V}H)$, it might be
  enough to truncate $\tilde V$ at a suitable polynomial (or
  perturbative, whatever is the small parameter in the expansion)
  order $L$; the required order $L$ can be determined on the base of
  the error $|\exp(L_{\tilde V}H)-\exp(L_{\tilde V^{(L)}}H)|$, which
  can be apriori estimated by exploiting Proposition ~\ref{p.2} in
  Appendix A.

\end{enumerate}

\medskip
\subsection{Third result: numerical evidence and the Lie-series of $P$.}

\begin{figure}[h!]
    \centering
    \includegraphics[width=0.47\textwidth]{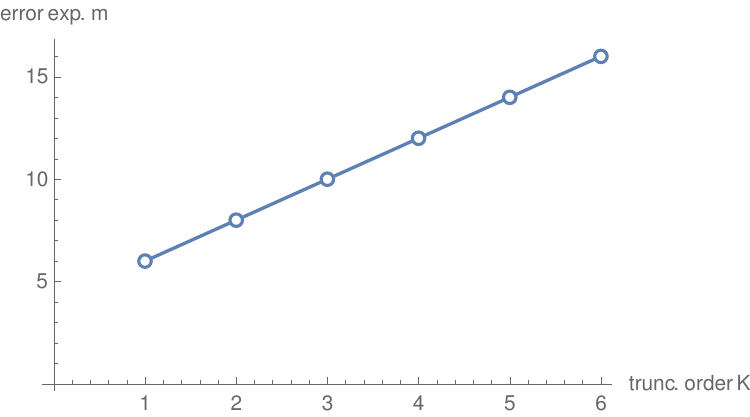}
    \hfill
    \includegraphics[width=0.47\textwidth]{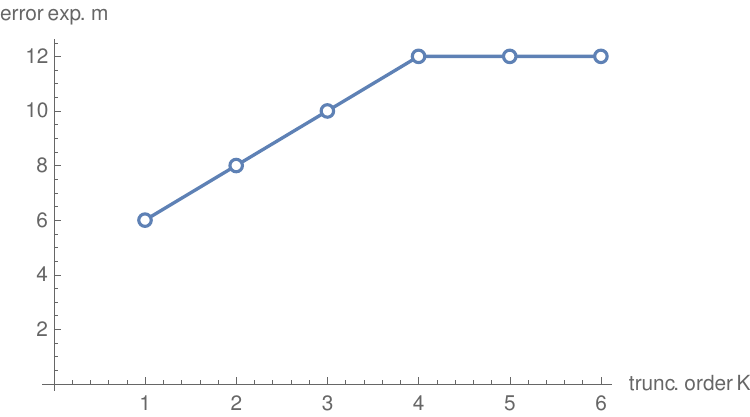}

    \caption{Order $m$ of the truncated Lie-series error
      $\Oscr(\rho^m)$ as a function of the truncation order $K$. Left
      panel: only truncation of the Lie-series is performed, hence the
      decay of the remainder is of the order
      $\Oscr(\rho^{2K+4})$. Right panel: truncation of the vector
      field $\tilde V^{(2L+1)}$ is added (with $L=4$), hence for
      $K\geq 4$ the error is kept constantly equal to
      $\Oscr(\rho^{12})$.}
    \label{fig:TrunLie}
\end{figure}

In order to apply the Lie-series method to transform the conserved
quantity $P$ given in \eqref{e.P_r}, and in general to transform any
function $f(q,p)$ through $\varphi^{-1}(x,y)$, we need the inverse
flow $\tilde\Phi^{-1}(1,x,y)$, which is indeed the flow of the
opposite field $-\tilde V(\tau,x,y)$. As a consequence, the
transformed quantity $P(\phi^{-1}(x,y))=P(\tilde\Phi^{-1}(1,x,y))$ can be
obtained from the Lie-series of $P$ along the field $-\tilde V$ and
has to coincide with \eqref{e.P_Dar}
\begin{equation*}
P(\tilde\Phi^{-1}(1,x,y)) = \exp(L_{-\tilde V}P)(1,x,y)=
\sum_{k\geq 0}\frac{(-1)^k}{k!}L^k_{\tilde V}P(1,x,y) = \frac{1}{2}\norm{(x,y)}^2\ .
\end{equation*}
This provides us with the possibility of numerically verifying the
convergence of the Lie series and estimating the approximation errors,
when the series is stopped at a certain order $K$ or when the vector
field is approximated with a suitable Taylor polynomial of order
$2L+1$.  Let $V^{(2L+1)}=\sum_{l=1}^L V_{t,2l+1}$ be the Taylor
polynomial at order $2L+1$ of $V_t$, and denote by $\exp_K(t L_{\tilde
  V}) = \sum_{k=0}^K\frac{t^k}{k!}L^k_{\tilde V}$ the Lie-series
truncated at order $K$; then the following result holds true:
\begin{proposition}
    \label{t.main.3}
    In a sufficiently small ball $B_\rho(0)\subset\RR^{2n}$ of the
    origin, the approximations of the Lie-series $\exp(L_{-\tilde
      V}P)(1,x,y)$ satisfy the following error estimates
    \begin{equation}
    \label{e.both.trunc.X}
        \begin{aligned}
        \sup_{(x,y)\in B_\rho}\Big|\frac12\norm{(x,y)}^2 -
        \exp_K(L_{-\tilde V}P)(1,x,y)\Big| &\leq C_1
        \nu^{K+1}\rho^{2K+4}\\ \sup_{(x,y)\in
          B_\rho}\Big|\frac12\norm{(x,y)}^2 - \exp_K(L_{-\tilde
          V^{(2L+1)}}P)(1,x,y)\Big| &\leq C_2
        \nu^{M+1}\rho^{2M+4}\ ,
        \end{aligned}
    \end{equation}
    with $M=\min\{L,K\}$ and suitable positive constants $C_{1,2}$.
\end{proposition}

Figure \ref{fig:TrunLie} shows the numerical
computation\footnote{Performed with Mathematica Release 12} of the
above errors for $K=1,\ldots,6$ and (in the right panel) $L=4$. The
exponent $m$ of the leading polynomial term in the error is plotted
against the truncation $K$. The left panel indicates a geometric
convergence of $\exp(L_{\tilde V}^{(K)}P)(1,x,y)$ to \eqref{e.P_Dar}
as $K$ is increased, in agreement with the first estimate in
\eqref{e.both.trunc.X}; the right panel shows that for $K\geq 4$ the
error is stabilized at values $\Oscr(\rho^{12})$, in agreement with
the second estimate in \eqref{e.both.trunc.X}, since $\min\{L,K\}=4$
for such values of $K$.


\section{Proofs of Theorems}
\label{s:proofs}

Given $x\in\RR^n$, we define the polydisk
$D_\rho=\bigotimes_{j=1,\ldots,n}\Bscr_\rho(x_j)\subset \CC^n$ as the
product of $n$-copies of the complex disk $\Bscr_\rho(x_j)$ with
radius $\rho\leq 1$ centered at the elements $x_j$. For $T>1$ and
$t\in \Gscr=(-T,T)$, we also define the extended complex domain
$\Gscr_{\delta}=\bigcup_{t\in\Gscr} \Bscr_\delta(t)\subset\CC$. Let
$\tilde X$ be a time dependent and analytic vector field $\tilde
X(t,x):\Gscr_\delta\times D_\rho\to\RR^{n+1}$, and let $f$ be a
real-valued function $f(t,x):\Gscr_\delta\times D_\rho\to\RR$ which is
analytic in the same domain, then we introduce the following norms
\begin{equation*}
  \normsup{f}{\delta,\rho}:=
  \sup_{(\zeta,z)\in\Gscr_\delta\times D_\rho}|f(\zeta,z)|
  \qquad\qquad
  \normsup{\tilde X}{\delta,\rho}:=
  \sup_{(\zeta,z)\in\Gscr_\delta\times D_\rho}|X(\zeta,z)|\ ,
\end{equation*}
where $|\tilde X(\zeta,z)|$ is the norm of a complex vector in
$\CC^{n+1}$. Clearly, if $f$ or $\tilde X$ depend only on the phase
space variable $z$, we will consider only $\normsup{f}{\rho}$ or
$\normsup{\tilde X}{\rho}$.  Moreover, we use the symbol $\prec$ to
compare to quantities, like the norms of two functions (or vector
fields), modulo a numerical constant
\begin{equation}
    \label{e.prec.symb}
    A\prec B
    \qquad\Longleftrightarrow\qquad
    \exists C>0 \quad\textrm{s.t.}\quad
    A\leq C B\ .
\end{equation}

\subsection{Proof of Theorem \ref{t.main.1}}

We first have to observe that \eqref{e.a} represents the integration
of the closed 2-form $-\eta=\omega_0-\omega_1$, where
$\omega_1=dq\wedge dp$ and the unknown $a(q,p)$ is the potential
vector of a 1-form $\alpha$ such that $d\alpha=-\eta$. Hence the
problem reduces to find $\alpha_{0}$ and $\alpha_1$ such that
\begin{equation*}
  d\alpha_0=\omega_0\qquad\qquad
  d\alpha_1=\omega_1\qquad\Rightarrow\qquad\alpha=\alpha_0-\alpha_1\ .
\end{equation*}
We notice that the two 2-forms already coincide at the origin,
$\omega_0(0,0)=\omega_1$, hence no preliminary linear transformation
is required.  We look for a potential of $\omega_0$. By passing to
action-angle like coordinates
$$
\begin{cases}
  q = \sqrt { A } \cos( \theta ) \\ p = - \sqrt { A } \sin( \theta )
\end{cases}
\qquad
\begin{cases}
  A=q^{2}+p^{2} \\ \theta=-\arctan\left (\frac{p}{q} \right )
\end{cases}
$$
one gets
$$
\quad d q \wedge d p=\frac{1}{2} d \theta \wedge d A\ .
$$
We can rewrite $\omega_0$ as
$$
\omega_0=\frac{1}{2(\nu A+1)}d\theta \wedge dA
$$
whose potential can be chosen as a $d\theta$-form
$\alpha_0=-\frac{1}{2\nu}\ln(1+\nu A)d\theta$, or in cartesian
coordinates $(q,p)$
$$
\alpha_0=\frac{\ln(1+\nu\norm{(q,p)}^2)}{2\nu\norm{(q,p)}^2}
\left (-pdq+qdp \right)\ .
$$
The main point in the solution of \eqref{e.a} is to choose the
potential $\alpha_1$ of $\omega_1$ as a $d\theta$-form as $\alpha_0$,
namely $\alpha_1=-\frac12Ad\theta$, or in cartesian coordinates
$\alpha_1=\frac{1}{2}\tond{-pdq+qdp}$, so that
$$
\alpha_0-\alpha_1=\frac{1}{2} \left [
  \frac{\ln{(1+\nu\norm{(q,p)}^2)}}{\nu\norm{(q,p)}^2}-1\right ](-pdq+qdp)\ ,
$$
which implies a solution $a(q,p)$ of \eqref{e.a} of the form
\begin{equation*}
  a(q,p) = \frac{1}{2} \left [
    \frac{\ln{(1+\nu\norm{(q,p)}^2)}}{\nu\norm{(q,p)}^2}-1\right ]
  \begin{pmatrix}
-p\\q
  \end{pmatrix}\ .
\end{equation*}
For $\Omega^{-\top}_t$ is given by $$\Omega^{-\top}_t=g(t,\nu\norm{(q,p)}^2) J \qquad\qquad
g= \frac{(1+\nu\norm{(q,p)}^2)}{1+t\nu\norm{(q,p)}^2}\ ,
$$
the vector field $V_t = \Omega^{-\top}_t a(q,p)$ reads
\begin{equation*}
  V_t=\chi(t,q,p) \begin{pmatrix}
  q\\p
  \end{pmatrix}\ ,\qquad\qquad\chi = \frac{1+\nu\norm{(q,p)}^2}{2(1+t\nu\norm{(q,p)}^2)}
  \quadr{\frac{\ln(1+\nu\norm{(q,p)}^2)}{\nu\norm{(q,p)}^2}-1} \ ,
\end{equation*}
which is \eqref{e.chi}.

\begin{lemma}
  \label{l:rot_sym}
  Given $T>1$, the vector field $V_t$ is contracting and analytic in
  $\Gscr_\delta\times D_\rho$ for $\delta\leq\frac1{\nu\rho^2}-T$ and
  $\rho\leq\rho_*:=\frac1{\sqrt{2T\nu}}$. Moreover:
  \begin{enumerate}

  \item it leaves the origin $O$ fixed and for $\norm{(q,p)}$ small
    enough it admits the time-independent asymptotic expansion
    $$
        V_t\sim V_{t,3}=-\frac14\nu\norm{(q,p)}^2
        \begin{pmatrix}
        q\\p
        \end{pmatrix}\ ;
    $$
    
  \item it is symmetric under the action of the rotation group
  \begin{displaymath}
    R(s)=
    \begin{pmatrix}
      \cos(s) & \sin(s)
      \\ -\sin(s) & \cos(s)
    \end{pmatrix}\ ;
  \end{displaymath}
  
  \item it satisfies the estimate
  \begin{equation*}
      \normsup{V_t}{T,\rho}\prec \nu\rho^3\ .
  \end{equation*}
  \end{enumerate}
\end{lemma}

\proof The field $V_t$ is clearly decomposed in a coefficient which
depends only on the norm $(q^2+p^2)$ and a radial direction $(q,p)$,
hence it is invariant under the group action of $R(s)$. Indeed $V_t$
commutes with $(p,-q)$, the generator of $R(s)$. Since
$\ln(1+\nu\norm{(q,p)}^2)<\nu\norm{(q,p)}^2$ for any $(q,p)\neq O$,
the flow is contracting in the future and the origin is the only
equilibrium of the dynamical system defined by $V_t$. The asymptotic
expansion $V_t\sim V_{t,3}$ is immediately derived from the Taylor
expansion with respect to the phase space variables $(q,p)$.  We
rewrite $\chi=\chi_1\chi_2$ with
\begin{equation*}
    \chi_1 = \frac{1+\nu\norm{(q,p)}^2}{2(1+t\nu\norm{(q,p)}^2)}\ ,
    \qquad\qquad\chi_2 =
    \frac{\ln(1+\nu\norm{(q,p)}^2)}{\nu\norm{(q,p)}^2}-1\ .
\end{equation*}
The second factor $\chi_2$ is analytic in $D_\rho$ with
$\rho<\frac1{\sqrt\nu}$; the first factor is analytic in polydisks
$B_\delta\times D_\rho$ where $\delta$, radius of the disks
$B_\delta(\zeta)$ for $\zeta\in (-T,T)$, satisfies
\begin{equation*}
    \delta =
    \inf_{|\zeta|<T}\Big|\frac1{\nu\norm{(q,p)}^2}+\zeta\Big|=\frac1{\nu\rho^2}-T\ ,
\end{equation*}
with $T\leq \frac1{\nu\rho^2}-T$ due to the condition
$\rho\leq\rho^*$. Hence we are allowed to take $\delta=T$ in the
estimate of $V_t$: the factor $\chi_1$ can be uniformly bounded by a
constant in the given domain, while the factor $\chi_2$ and the radial
direction provide the cubic dependence on $\rho$.  \qed

The above Lemma implies that its flow $\Phi^1$ is a (contracting in
the future) nonlinear analytic deformation of the identity map, for
any $|t|\leq 1$, provided $\rho$ is small enough. Indeed, by its
definition in terms of Lie-series\footnote{Alternatively, one can
  construct a series of analytic approximating solutions which are
  uniformly convergent in $B_\rho$ to the solution, then also the
  solution has to be analytic.}, in Lemma \eqref{l:rot_sym} we can
take $\delta=T=\frac{1}{2\nu\rho^2}$ (which means $\rho=\rho^*$) and
$d_1=d_2=d$ so that
\begin{equation*}
    \Gamma\prec\frac{\nu\rho^2}{d}<\Gamma^*
\end{equation*}
holds true for sufficiently small $\rho$; the asymptotic expansion
\eqref{e.phi.exp} follows immediately from the local behavior of
$V_t$
\begin{equation*}
    \Phi^1(q,p)\sim 
    \tond{1+V_t(1,q,p)} \begin{pmatrix}
        q\\p
    \end{pmatrix} \sim 
    \tond{1-\frac14\nu\norm{(q,p)}^2}\begin{pmatrix}
        q\\p
    \end{pmatrix}\ .
\end{equation*}

We finally have to show that $\Phi^1(q,p)$ is the inverse of
\eqref{e.D_trsf}.  We can assume the time-one-flow
$\varphi(q,p)=\Phi^1(q,p)$ to be a radial and close to the identity
transformation having the form
\begin{equation*}
  \Phi^1(q,p) = \xi(\sqrt\nu\norm{(q,p)})\begin{pmatrix}
    q\\p
    \end{pmatrix}\qquad\qquad \xi(0)=1\ ,
\end{equation*}
with $\xi$ being analytic in the norm $\norm{(q,p)}$; the same can
be assumed also for the inverse transformation $\varphi^{-1}(x,y)$
\begin{equation*}
  \varphi^{-1}(x,y) = \sigma(\sqrt\nu\norm{(x,y)})\begin{pmatrix}
    x\\y
    \end{pmatrix}\qquad\qquad \sigma(0)=1\ .
\end{equation*}
By imposing for $\varphi^{-1}(x,y)$ the condition of being a Darboux
transformation $(\varphi^{-1})^*\omega_0 = \omega_1$, we get the
following equation for $\sigma$
\begin{equation*}
\sigma'\sigma \varrho+\sigma^2=1+\varrho^2\sigma^2\qquad \varrho=\sqrt\nu\norm{(x,y)}\,
\end{equation*}
which becomes, by introducing the more suitable variable
$h(\varrho)=\sigma^2(\varrho)$, a linear and non-homogeneous equation of the form
\begin{equation}
  \label{e.psi}
h'+\frac2\varrho(1-\varrho^2)h -\frac2\varrho=0\ .
\end{equation}
The unique analytic solution of \eqref{e.psi} is given by
\begin{displaymath}
h(\varrho) = \frac1{\varrho^2}\tond{e^{\varrho^2}-1}\ ,
\end{displaymath}
which gives $\sigma(x,y)$ the expression in \eqref{e.D_trsf}.

\subsection{Proof of Theorem \ref{t.main.2}}

In order to prove the normal form statement and the bound on
$\Rscr^{(1)}$, we make use of three different levels of
approximation. As in the previous proof, we set
$\delta=T=\frac1{2\nu\rho^2}$ and we assume $\rho$ small enough to
ensure $T>1$. First expand the vector field $\tilde V$
\begin{equation*}
    \exp(L_{-\tilde V} H) = \exp(L_{-\tilde V_3} H) + \Rscr_1\ ,
\end{equation*}
where $\Rscr_1$ can be bounded using Proposition \eqref{p.3} with
$\Gamma_3 = \frac1{d\rho}\normsup{V-V_t}{T,\rho}\prec \nu^2\rho^4$
\begin{equation*}
    \normsup{\Rscr_1}{(1-d)T,(1-d)\rho}\prec
    \Gamma_3\normsup{H}{T,\rho}\prec \nu^3\rho^8 + \eps\nu^2\rho^6\ .
\end{equation*}
As a second step, we truncate the Lie series of $\tilde V_3$ at order $K=1$
\begin{equation*}
    \exp(L_{-\tilde V_3} H) = \exp_1(L_{-\tilde V_3} H) + \Rscr_2 \ ,
\end{equation*}
where
\begin{equation*}
    \exp_1(L_{-\tilde V_3} H) = H-L_{\tilde V_3} H\ ,
\end{equation*}
and $\Rscr_2$ can be bounded using Proposition \eqref{p.2} with
$\Gamma=\frac1d\tond{2\nu\rho^2 + \normsup{\tilde V}{T,\rho}}\prec
\nu\rho^2$
\begin{equation*}
    \normsup{\Rscr_2}{(1-d)T,(1-d)\rho}\prec\Gamma^2\normsup{H}{T,\rho}\prec
    \nu^3\rho^8+\eps\nu^2\rho^6\ .
\end{equation*}
Last and easiest step consists in a Taylor expansion of $H_0$ in
$H-L_{\tilde V_3} H$ which gives
\begin{equation*}
    H-L_{\tilde V_3} H = H_{0,4} + H_{1,2}  - L_{V_3}H_{0,4} - L_{V_3}H_{1,2} + \Rscr_3\ ,
\end{equation*}
with
\begin{equation*}
    \normsup{\Rscr_3}{\rho} \prec \eps\nu^2\rho^6 + \nu^3\rho^8\ .
\end{equation*}
Since $d<1$ is arbitrary, $\rho$ is assumed to be small enough and the
three contributions $\Rscr_j$ to the remainder are of the same order,
the estimate follows. The estimates for $\Rscr^{(0)}$ follows by minor
variations; in the AL case just remember that the cubic nonlinearity
is one of the leading terms in the remainder.

In order to prove the different bounds on the closeness between the
AL/Salerno model and the normal forms, one has to apply Cauchy
estimates \eqref{e.Cauchy} to get estimates on the vector fields
$X_{\Rscr^{(0,1)}}$, starting from the bounds on the remainders
\begin{equation}
\label{e.Cauchy.X}
  \normsup{X_{\Rscr}}{(1-d)\rho}\leq\frac1{d\rho}\normsup{\Rscr}{\rho}\ ,\qquad\qquad
  \normsup{X_{\Rscr}}{(1-d)\rho} = \max_{j=1,\ldots,2n}\normsup{X_{\Rscr,j}}{(1-d)\rho}\ .
\end{equation}
Hence from \eqref{e.Cauchy.X} one can obtain
\begin{equation*}
  \normsup{X_{\Rscr^{(1)}}}{(1-d)\rho}\prec \nu^2\rho^5(\rho^2+\eps)\ ,
\end{equation*}
for the cubic-quintic dNLS-like normal form and
\begin{equation*}
  \normsup{X_{\Rscr^{(0)}}}{(1-d)\rho}\prec \nu^2\rho^3(\rho^2+\eps)\ ,\qquad\qquad
    \normsup{X_{\Rscr^{(0)}}}{(1-d)\rho}\prec \nu^2\rho^3(1+\eps)\ ,
\end{equation*}
for the standard dNLS normal form, in the Salerno and AL cases
respectively. Then all the estimates follow from Gronwall Lemma and
from the conservation of the norm $\norm{x(t),y(x)}^2$ along the
different Hamiltonian flows (so that all the orbits belong to the
initial ball $B_\rho(0)$ for infinite times). We can sketch the
procedure as follows: we have to compare the dynamics of
$H=Z^{(0)}+\Rscr^{(0)}$ with the one of the normal form $Z^{(0)}$
\begin{equation*}
\dot z = X_H(z)\qquad\qquad \dot \zeta = X_{Z^{(0)}}(\zeta) = X_H(\zeta)-X_{\Rscr^{(0)}}(\zeta)\ ,
\end{equation*}
so we introduce the error $\delta(t)=z(t)-\zeta(t)$ which solves
\begin{equation*}
\dot \delta = [X_H(z) - X_H(\zeta)] + X_{\Rscr^{(0)}}(\zeta) = [X_H(\zeta+\delta)
  - X_H(\zeta)] + X_{\Rscr^{(0)}}(\zeta) \ .
\end{equation*}
Hence we can derive the differential inequality
\begin{equation*}
\norm{\dot \delta} \leq \norm{X_H(\eta)}\norm{\delta} +
\norm{X_{\Rscr^{(0)}}(\zeta)} \ ,\qquad\qquad \eta =
\zeta+s\delta\ ,\qquad s\in[0,1]\ ;
\end{equation*}
then Gronwall estimate can be applied, once we provide estimates for
$\norm{X_H(\eta)}$ and $\norm{X_{\Rscr^{(0)}}(\zeta)}$, for
$\zeta,\,\eta$ belonging to a polydisk of small radius.

\subsection{Proof of Proposition \ref{t.main.3}}

To prove both the estimates, we have first to set (as in the previous
proof) $\delta=T=\frac1{2\nu\rho^2}$ and assume $\rho$ small enough,
so that $T>1$ and we can evaluate the Lie-series at $\tau=1$, lying
inside the domain $\Gscr_\delta$. The first of \eqref{e.both.trunc.X}
is a consequence of \eqref{eq.est.trunc.K} with
$\normsup{f}{\delta,\rho}=\normsup{P}{\delta,\rho}\sim\rho^2$ and
\begin{equation*}
    \Gamma\prec\frac{\nu\rho^2}{d}\qquad\Rightarrow\qquad 
    (e\Gamma)^{K+1}\normsup{f}{\delta,\rho}\prec \nu^{K+1}\rho^{2K+4}\ .
\end{equation*}
The second of \eqref{e.both.trunc.X} can be derived combining the
previous estimate with \eqref{e.trunc.X}, where $\Gamma_3$, defined
in \eqref{e.gammas}, in this case fulfills
\begin{equation*}
    \Gamma_3\prec\frac1{\rho}\normsup{Y}{\delta,\rho} = 
    \frac1{\rho}\normsup{V_t-V_t^{(2L+1)}}{\delta,\rho} \sim \nu^{L+1}\rho^{2L+2}\ ,
\end{equation*}
since the leading term of the remainder $V_t-V_t^{(2L+1)}$ is of order $2L+3$.


\section{Conclusions}
\label{s:concl}

In this manuscript we have focused on the constructive aspects of
Moser's proof of Darboux's Theorem, with the aim of a deeper
understanding of the standardization procedure, in particular for the
AL and Salerno model.

In general, Darboux's change of coordinates is local, and linear
coordinates can always be chosen so that it represents a small
perturbation of the identity map; hence a successful strategy is to
combine the polynomial approximation of the vector field $V$ with the
use of truncated Lie-series (or even Lie-transform), in order to
compute leading order terms of the transformed Hamiltonian $H(\Psi)$,
{\em without any need to derive a complete explicit expression for
  $\Psi$ or for $V$ itself}. This approximation of the transformed
Hamiltonian might be enough for the subsequent investigation of its
dynamical features by means of perturbation techniques (for example,
by normal form methods). For what concerns the specific case of the
Salerno model, such a normal form can be used as a starting point for
perturbation schemes which require the use of the standard Poisson
brackets: for example, one can start to explore existence and
stability of localized solutions, such as multi-breathers,
quasi-periodic breathers, low dimensional tori (see
\cite{DanSPP22,You98,Yuan02,JohA97,GengVY08}).

Outside the field of nonlinear lattices, a classical example that it
is worth mentioning is that of the Lotka-Volterra system and its
higher dimensional generalizations (Lotka-Volterra systems). It is
well known that, for $(x,y)\in\RR^2$ the predator-prey system $\dot
x=\alpha x - \beta xy$, $\dot y=-\gamma y + \delta xy$ (usually with
all the parameters taken positive) admits the constant of motion
$H(x,y)=\beta y - \alpha \ln y + \delta x - \gamma \ln x$. In the
original variables the system is not Hamiltonian with respect to the
standar Poisson structures, but it is so with the non
standard\footnote{Sometimes called log-canonical.}  Poisson brackets
$\Poi{x}{y}=xy$. Equivalently it is possible to use the change of
variables $\xi=\ln x$, $\eta=\ln y$ to obtain an Hamiltonian form with
the standard symplectic structure. In~\cite{SS94} it is discussed the
effectiveness of a particular numerical integration scheme by showing
that it turns out to be symplectic with respect to the non standard
Poisson structure. A suitable generalization of the above mentioned
non standard structure is also used in the higher dimensional
extensions of LV models: the possibility to view also those systems as
Hamiltonian open the way for the investigation of their integrability
(see~\cite{CHK19} for a recent work in that direction).


\appendix

\section{Lie series in the extended phase space.}
\label{s:Lie}

In this Section we present some analytical result about Lie-series in
the non autonomous case (see for example the Appendix of
\cite{GioLS15} for similar estimates in the autonomous case and
\cite{GioZ92} in the non autonomous Hamiltonian context).  In order to
estimate the Lie series of a given function $f$
\begin{equation*}
\exp(t L_{\tilde X} f)(\zeta,z)=\sum_{k\geq 0}\frac1{k!}t^k L^k_{\tilde X} f (\zeta,z)    
\end{equation*} 
already defined in \eqref{e.Lie.op}, it is necessary to provide an
upper bound to the Lie derivative $L_{\tilde X} f$ in the shrinked
domain $\Gscr_{(1-d_1)\delta}\times D_{(1-d_2)\rho}$, according to the
usual Cauchy inequality valid for analytic functions (in several
complex variables)
\begin{equation}
  \label{e.Cauchy}
    \normsup{\derp{f}{\zeta}}{(1-d_1)\delta,(1-d_2)\rho} \leq
    \frac1{d_1\delta}\normsup{f}{\delta,(1-d_2)\rho}\ ,\qquad
    \normsup{\derp{f}{z_j}}{(1-d_1)\delta,(1-d_2)\rho} \leq
    \frac1{d_2\rho}\normsup{f}{(1-d_1)\delta,\rho}\ ,
\end{equation}
which is a consequence of the (multidimensional) Cauchy
formula\footnote{One can differently consider $\derp{f}{u_j}(u) =
  \derp{f(u+e_jz)}{z}(0)$ and use the one dimensional version of the
  Cauchy formula.}
\begin{equation*}
    \derp{^kf}{u^k}(u) =
    \frac{k!}{2\pi\im}\int_{T_{\delta,\rho}}\frac{f(v)}{(v-u)^{k+1}}dv\ ,\qquad
    u=(\zeta,z)\ ,
\end{equation*}
where $k=(k_0,k_1,\ldots,k_n)\in\mathbb{Z}^{n+1}$ is a derivative
multi index and $T_{a,b}=\partial B_{\delta}(\zeta)\times\partial
B_{\rho}(z_1)\times\ldots\times B_{\rho}(z_n)$ is the
$n+1$-dimensional torus given by the product of the boundaries of all
the distinguished disks (distinguished boundary). Hence, if $f$ is
analytic in $D_\rho$, the same Cauchy formula allows to bound the
$k$-derivative of $f$ (with respect to $z$) in $D_{\rho-\delta}$
\begin{equation*}
    \normsup{\derp{^lf}{z^l}}{\rho-\delta}\leq\frac{l_1!\ldots
      l_n!}{\delta^{k}}\normsup{f}{\rho}\ ,\qquad |l|=k\ .
\end{equation*}
Consider $f$, analytic on $\Gscr_\delta\times D_1$, as an analytic
function on $D_\rho\subset D_1$, and Taylor expand $f$ with respect to
$z\in D_\rho$
\begin{equation*}
    f(\zeta,z) = T_k(\zeta,z) + R_{k+1}(\zeta,z)\ ,\qquad\qquad
    R_{k+1}(\zeta,z)=\sum_{|l|= k+1}\frac1{l!}
    f^{(l)}(\zeta,cz)z^l\ ,\qquad c\in(0,1)\ .
\end{equation*}
Then the following estimate provide the exponential decay of the reminder ot the truncated Taylor series
\begin{lemma}
    \label{l.exp.decay}
    Let $\rho<\frac12$, then there exists $E=E(k,n)$ such that
    \begin{equation*}
        \normsup{R_{k+1}}{\delta,\rho}\leq (2\rho)^{k+1}\normsup{f}{1}E\ .
    \end{equation*}
\end{lemma}
\begin{proof}
It is enough to exploit the above bound of the derivative with respect
to $z$ and the combinatorial formula of the number of partial
derivatives of given order $k+1$. Indeed from
\begin{equation*}
    |R_{k+1}(\zeta,z)|\leq \sum_{|l|= k+1}\frac1{l!} |f^{(l)}(\zeta,cz)| \rho^{k+1}\ ,
\end{equation*}
since $z\in D_{\rho}$ implies $cz\in D_{c\rho}\subset D_\rho$, we have
\begin{equation*}
    \normsup{R_{k+1}}{\delta,\rho}\leq \rho^{k+1} \sum_{|l|=
      k+1}\frac1{l!} \normsup{f^{(l)}}{\delta,\rho} \ ,
\end{equation*}
with
\begin{equation*}
    \normsup{f^{(l)}}{\delta,\rho}\leq
    \frac{l!}{(1-\rho)^{k+1}}\normsup{f}{1}<
    2^{k+1}l!\normsup{f}{1}\ .
\end{equation*}
Hence
\begin{equation*}
    \normsup{R_{k+1}}{\delta,\rho}\leq (2\rho^{k+1}) \normsup{f}{1}
    \tond{\sum_{|l|= k+1}} = (2\rho^{k+1}) \normsup{f}{1} E \ ,
\end{equation*}
where $E=C^*_{n,k+1} = \binom{n+k}{k+1}$.
\end{proof}
\begin{lemma}
\label{lem.1}
Let $\tilde X=(1,X)$ with $X$ analytic in $\Gscr_\delta \times D_\rho$
and $f$ analytic in $\Gscr_\delta\times D_\rho$, and let $0<d_j<1$,
then
    \begin{equation}
        \label{e.lie.est.1}
        \normsup{L_{\tilde X} f}{(1-d_1)\delta,(1-d_2)\rho}\leq
        \Gamma\normsup{f}{\delta,\rho}\qquad\qquad
        \Gamma=\frac1{d_1\delta} +
        \frac{\normsup{X}{\delta,\rho}}{d_2\rho}\ .
    \end{equation}
\end{lemma}

\begin{proof}
    Notice that the Lie differential operator $L_{\tilde X}$ acts on
    $f$ as
    \begin{equation*}
        L_{\tilde X} f = \partial_\zeta f+ \inter{X,\nabla_z f}\ ,
    \end{equation*}
    hence from the previous Cauchy estimate and the usual bound on the scalar product we get \eqref{e.lie.est.1}.
\end{proof}

\begin{remark}
    The basic estimate \eqref{e.lie.est.1} and the ones which follow
    are equivalent to the ones due to Gr\"obner in \cite{Grob73},
    obtained originally with the classical methods of majorants due to
    Cauchy.
\end{remark}
    
\begin{remark}
    If $f$ and $X$ depend only on $z\in D_\rho$, then formula
    \eqref{e.lie.est.1} becomes the usual estimate
    \begin{equation*}
        \normsup{L_{X} f}{(1-d)\rho}\leq
        \frac{1}{d\rho}\normsup{X}{\rho}\normsup{f}{\rho}\ .
    \end{equation*}
\end{remark}

Next Lemma provides the main estimate when dealing with the
convergence of the Lie series and related results
\begin{lemma}
\label{lem.2}
    Let $\tilde X=(1,X)$  and $f$ as in Lemma \ref{lem.1}, and let $0<d_j<1$, then
    \begin{equation}
        \label{e.lie.est.k}
        \normsup{L^k_{\tilde X} f}{(1-d_1)\delta,(1-d_2)\rho}\leq
        \frac{k!}{e}\tond{\Gamma e}^k \normsup{f}{\delta,\rho}\ .
    \end{equation}
\end{lemma}

\begin{proof}
The result follows by repeating formula \eqref{e.lie.est.1} for
$k$-times to each Lie derivative, starting from $L^k_{\tilde X} f =
L_{\tilde X} \tond{L^{k-1}_{\tilde X}}f$, by reducing each time the
domain of the supremum norm by the factors $d'_j=\frac1k d_j$. Then,
to conclude, one has to use the basic inequality $k^k\leq k! e^{k-1}$.
\end{proof}

The above estimate \eqref{e.lie.est.k} can be generalized in the
following way
\begin{lemma}
\label{lem.3}
    Let $\tilde X_j=(1,X_j),\,j=1,\ldots,k$, be a sequence of vector
    fields with $X_j$ analytic in $\Gscr_\delta \times D_\rho$ and $f$
    analytic in $\Gscr_\delta\times D_\rho$, and let $0<d_j<1$, then
    \begin{equation}
        \label{e.lie.est.gen}
        \normsup{L_{\tilde X_k}\ldots L_{\tilde X_1}
          f}{(1-d_1)\delta,(1-d_2)\rho}\leq \frac{k!
          e^k}{e}\tond{\prod_{j=1}^k \Gamma_j}
        \normsup{f}{\delta,\rho}\qquad\qquad \Gamma_j =
        \frac1{d_1\delta} +
        \frac{\normsup{X_j}{\delta,\rho}}{d_2\rho}\ .
    \end{equation}
\end{lemma}

Next Lemma exploits the previous estimates to get total convergence of
the Lie series on a smaller domain $\Gscr_\delta'\times D_\rho'\subset
\Gscr_\delta\times D_\rho$:
\begin{lemma}
\label{l.tot.conv}
    Let $\tilde X$ and $f$ be as in Lemma \ref{lem.1}. Then for any
    $0<d_j<1$ the Lie series $\exp(t L_{\tilde X} f)$ is totally
    convergent in $\Gscr_{(1-d_1)\delta}\times D_{(1-d_2)\rho}$ for
    times $|t|<\frac1{e\Gamma}$.
\end{lemma}

\begin{proof}
By using \eqref{e.lie.est.gen} one obtains
    \begin{equation*}
        \normsup{\exp(t L_{\tilde X}
          f)}{(1-d_1)\delta,(1-d_2)\rho}\leq \sum_{k\geq 0}
        \frac1{k!}|t|^k\normsup{L^k_{\tilde X}
          f}{(1-d_1)\delta,(1-d_2)\rho}\leq
        \frac{\normsup{f}{\delta,\rho}}e\sum_{k\geq 0}(|t|\Gamma
        e)^k<\infty\ ,
    \end{equation*}
    for times $|t|\Gamma e<1$.
\end{proof}

In the following we focus on the Lie series at times $t=\pm 1$; indeed
we are mainly interested in the transformation $\varphi(z):=
\graff{\varphi_j(z)}_{j=1,\ldots,n}$, with
$\varphi_j(z)=\exp(L_{\tilde X} \Id_j)(0,z)$, and in its inverse
$\varphi^{-1}(z)$. The following result holds true:
\begin{proposition}
\label{prop.1}
    Let $\tilde X$ be as in Lemma \ref{lem.1}. Then for any
    $j=1,\ldots,n$ the Lie series $\exp(t L_{\tilde X} \Id_j)(0,z)$ is
    totally convergent in $z\in D_{(1-d)\rho}$, for times
    $|t|<\frac1{\Gamma e}$ and $0<d<1$. Moreover, if
    \begin{equation*}
        \Gamma<\Gamma^*=\frac{e}{1+e^2}<\frac1e\ ,
    \end{equation*}
    then the transformations $\varphi^{\pm 1}(z)$ are well defined as
    Lie-series and, for $0<d<\frac12$, satisfy the inclusion
    \begin{equation*}
        D_{(1-2d)\rho}\subset \varphi^{\pm 1}(D_{(1-d)\rho})\subset
        D_\rho\ .
    \end{equation*}
\end{proposition}

\begin{proof}
    For $z\in D_{(1-d_2)\rho}$ we have
    \begin{equation*}
        |\exp(t L_{\tilde X} \Id_j)(0,z)|\leq \normsup{\exp(t
          L_{\tilde X} \Id_j)}{(1-d_1)\delta,(1-d_2)\rho}\leq
        \frac{\normsup{\Id_j}{\rho}}e\sum_{k\geq 0}(|t|\Gamma
        e)^k<\infty\ .
    \end{equation*}
    Moreover, condition $\Gamma<\Gamma^*$ ensures
    $\frac{1}{e\Gamma}>1$, so that $t=\pm1$ makes sense in the
    Lie-series. The deformation of the domain can be bounded by taking
    the Lie series from $k\geq 1$
    \begin{equation*}
        \normsup{\exp(L_{\tilde X}
          \Id_j)-\Id_j}{(1-d_1)\delta,(1-d_2)\rho}\leq\sum_{k\geq 1}
        \frac1{k!}\normsup{L^k_{\tilde X}
          \Id_j}{(1-d_1)\delta,(1-d_2)\rho} = \sum_{k\geq 1}
        \frac1{k!}\normsup{L^{k-1}_{\tilde X}
          X_j}{(1-d_1)\delta,(1-d_2)\rho}\ ;
    \end{equation*}
    then
    \begin{align*}
        \normsup{L^{k-1}_{\tilde X} X_j}{(1-d_1)\delta,(1-d_2)\rho}
        &\leq \frac{(k-1)!}e (e\Gamma)^{k-1}\normsup{X_j}{\delta,\rho}
        = (d_2\rho)\frac{(k-1)!}e (e\Gamma)^{k-1}
        \frac{\normsup{X_j}{\delta,\rho}}{d_2\rho}\leq \\ &\leq
        (d_2\rho)\frac{(k-1)!}{e^2} (e\Gamma)^k\ .
    \end{align*}
    Hence by condition $\Gamma<\Gamma^*$ one has
    \begin{equation*}
        \sum_{k\geq 1} \frac1{k!}\normsup{L^k_{\tilde X}
          \Id_j}{(1-d_1)\delta,(1-d_2)\rho}\leq
        \frac{d_2\rho}{e^2}\tond{\frac{e\Gamma}{1-e\Gamma}}<d_2\rho\ .
    \end{equation*}
    The same estimate holds by replacing $(1-d_2)\rho$ with
    $(1-2d_2)\rho$ and $\rho$ with $(1-d_2)\rho$ respectively in the
    inequalities involving the Lie derivatives.
\end{proof}

The proof clearly shows that the size of $\normsup{X}{\delta,\rho}$
provides the leading deformation of $\varphi(z)$ with respect to the
identity
\begin{equation*}
    \normsup{\varphi_j(z) - z_j}{(1-d_2)\rho}\prec
    \normsup{X_j}{\delta,\rho}\ .
\end{equation*}

Two important issues to be addressed in perturbation theory are to
estimate the error when either the Lie series is truncated at some
finite order $K$ or the analytic vector field $\tilde X$ is
approximated with its Taylor polynomial, truncated at order $L$. To
provide standard estimates of this type is the goal of next
results. We first introduce the following notation: given two integers
$K,L\geq 1$ we denote by $\exp_K(L_{\tilde X}f) = \sum_{k=0}^K
L^k_{\tilde X}f$, the truncation of the Lie series at order $K$, and
by $\exp(L_{\tilde X^{(L)}}f) = \sum_{k\geq 0} L^k_{\tilde X^{(L)}}f$,
the Lie series of the truncated vector field $\tilde X^{(L)}$ at order
$L$. In the following, to simplify some estimates, we also make use of
the symbol $\prec$ defined in \eqref{e.prec.symb}.

\begin{proposition}
    \label{p.2}
    Let $\tilde X$ and $f$ as in Lemma \ref{lem.1},\,$0<d_j<1$ and
    $K\geq 1$. If $\Gamma<\Gamma^*$ then
    \begin{equation}
        \label{eq.est.trunc.K}
        \normsup{\exp(L_{\tilde X}f)-\exp_K(L_{\tilde X}f)}{(1-d_1)\delta,(1-d_2)\rho}
        \prec \tond{e\Gamma}^{K+1}\normsup{f}{\delta,\rho}\ .
    \end{equation}
\end{proposition}

\begin{proof}
    The result is a consequence of the estimate of the reminder
    \begin{align*}
        \normsup{\sum_{k\geq K+1} L^k_{\tilde
            X}f}{(1-d_1)\delta,(1-d_2)\rho} &\leq
        \frac1e\normsup{f}{\delta,\rho}\sum_{k\geq
          K+1}\tond{e\Gamma}^k = \frac1e
        \tond{e\Gamma}^{K+1}\normsup{f}{\delta,\rho} \tond{\sum_{k\geq
            0}\tond{e\Gamma}^k}\ .
    \end{align*}
\end{proof}

In order to study the error due to the truncation of a vector field,
we start considering the Lie series of two vector fields $\tilde
X_{1,2}$. Next Lemma allows to properly rewrite the difference of Lie
derivatives:

\begin{lemma}
\label{lem.4}
    Let $\tilde X_1$ and $\tilde X_2$ two vector fields as in Lemma
    \ref{lem.1}, and define $\tilde Y=\tilde X_1 - \tilde X_2 =
    (0,X_1-X_2)$. Then the following holds true
    \begin{equation}
        \label{e.X1.X2}
        \begin{aligned}
        L_{\tilde X_1} - L_{\tilde X_2} &= L_{\tilde Y}\ ,\\
        L^k_{\tilde X_1} - L^k_{\tilde X_2} &= L_{\tilde Y} L^{k-1}_{\tilde X_2} + 
        \sum_{l=1}^{k-1} L^{k-l}_{\tilde X_1}L_{\tilde Y} L^{l-1}_{\tilde X_2}\qquad k\geq 2\ ,\\
        \end{aligned}
    \end{equation}
    where obviously $L_{\tilde Y} = L_{Y}$.
\end{lemma}

\begin{proof}
    By induction. Formula \eqref{e.X1.X2} works for $k=1$, since by
    linearity we have $L_{\tilde X_1} - L_{\tilde X_2}=L_{\tilde Y}$,
    and for $k=2$, since
    \begin{equation*}
      L^2_{\tilde X_1} - L^2_{\tilde X_2} = L_{\tilde X_1}L_{\tilde
        X_2} + L_{\tilde X_1}L_{\tilde Y} - L^2_{\tilde X_2} =
      L_{\tilde X_1}L_{\tilde Y} +L_{\tilde Y}L_{\tilde X_2}\ .
    \end{equation*}
    We assume the above formula to hold for $k\geq 2$ and we show its
    validity for $k+1$. Indeed
    \begin{align*}
        L^{k+1}_{\tilde X_1} - L^{k+1}_{\tilde X_2} &= L_{\tilde
          X_1}L^k_{\tilde X_1} - L^{k+1}_{\tilde X_2} = L_{\tilde
          X_1}\quadr{L^k_{\tilde X_2} + L_{\tilde Y} L^{k-1}_{\tilde
            X_2} + \sum_{l=1}^{k-1} L^{k-l}_{\tilde X_1}L_{\tilde Y}
          L^{l-1}_{\tilde X_2}} - L^{k+1}_{\tilde X_2} = \\ &=
        L_{\tilde X_1} L^k_{\tilde X_2} + L_{\tilde X_1}L_{\tilde Y}
        L^{k-1}_{\tilde X_2} + \sum_{l=1}^{k-1} L^{k+1-l}_{\tilde
          X_1}L_{\tilde Y} L^{l-1}_{\tilde X_2}- L^{k+1}_{\tilde X_2}
        = \\ &= L_{\tilde Y} L^{k}_{\tilde X_2} + \sum_{l=1}^{k}
        L^{k+1-l}_{\tilde X_1}L_{\tilde Y} L^{l-1}_{\tilde X_2} \ .
    \end{align*}
\end{proof}

\begin{proposition}
    \label{p.3}
    Let $\tilde X_{1,2}$ and $\tilde Y$ as in Proposition \ref{p.2}
    and let $\Gamma_{j=0,1,2,3}$ be defined by
    \begin{equation}
        \label{e.gammas}
        \Gamma_1=\frac1{d_1\delta}+\frac{\normsup{X_1}{\delta,\rho}}{d_2\rho}\ ,\qquad
        \Gamma_2=\frac1{d_1\delta}+\frac{\normsup{X_2}{\delta,\rho}}{d_2\rho}\ ,\qquad
        \Gamma_3=\frac{\normsup{Y}{\delta,\rho}}{d_2\rho}\ ,\qquad \Gamma_0=\max\graff{\Gamma_1,\Gamma_2}\ .
    \end{equation}
    Then for any $f$ analytic in $\Gscr_\delta\times D_\rho$, if
    $e\Gamma_0<1$ we have
    \begin{equation}
        \label{e.diff.X1.X2}
        \normsup{ \exp(L_{\tilde X_1} f) - \exp(L_{\tilde X_2} f)}{(1-d_1)\delta,(1-d_2)\rho} \prec
        \Gamma_3 \normsup{f}{\delta,\rho}\ .
    \end{equation}
\end{proposition}

\begin{proof}
    We have to exploit Lemma \ref{lem.4} to estimate the difference of
    the Lie derivatives for $k\geq 2$
    \begin{align*}
        \normsup{ L^k_{\tilde X_1} f - L^k_{\tilde X_2}
          f}{(1-d_1)\delta,(1-d_2)\rho} &\leq \normsup{L_{\tilde Y}
          L^{k-1}_{\tilde X_2}f}{(1-d_1)\delta,(1-d_2)\rho} +
        \sum_{l=1}^{k-1}\normsup{L^{k-l}_{\tilde X_1}L_{\tilde Y}
          L^{l-1}_{\tilde X_2}f}{(1-d_1)\delta,(1-d_2)\rho} \leq
        \\ &\leq k!
        e^{k-1}\normsup{f}{\delta,\rho}\quadr{\Gamma_3\Gamma_2^{k-1} +
          \sum_{l=1}^{k-1} \Gamma_1^{k-l}\Gamma_3\Gamma_2^{l-1}} =
        \\ &=k! e^{k-1}\normsup{f}{\delta,\rho}\Gamma_3
        \quadr{\sum_{l=1}^{k}\Gamma_1^{k-l}\Gamma_2^{l-1}}\leq
        k!(k-1)\Gamma_3\tond{e\Gamma_0}^{k-1}
        \normsup{f}{\delta,\rho}\ ,
    \end{align*}
    while for $k=1$ one easily has
    \begin{equation*}
        \normsup{ L_{\tilde X_1} f - L_{\tilde X_2}
          f}{(1-d_1)\delta,(1-d_2)\rho} = \normsup{ L_{Y}
          f}{(1-d_1)\delta,(1-d_2)\rho}\leq
        \Gamma_3\normsup{f}{\delta,\rho}\ .
    \end{equation*}
    Hence we have
    \begin{equation*}
        \normsup{ \exp(L_{\tilde X_1} f) - \exp(L_{\tilde X_2}
          f)}{(1-d_1)\delta,(1-d_2)\rho} \leq
        \Gamma_3\normsup{f}{\delta,\rho}\tond{1+\sum_{k\geq
            1}k\tond{e\Gamma_0}^{k}}\prec
        \Gamma_3\normsup{f}{\delta,\rho} \ .
       \end{equation*}
    \end{proof}

The previous Proposition allows to estimate the error due to the
truncation of a vector field. Indeed, if $X_2=X^{(L)}$, namely the
truncation at order $L$ of a given field $X$, then $Y=X-X^{(L)}$ is
the remainder. Due to Lemma \ref{l.exp.decay} it holds also
$\normsup{Y}{\delta,\rho}\leq \normsup{X}{\delta,1}\rho^{L+1}E$, hence
\eqref{e.diff.X1.X2} takes the form
\begin{equation}
    \label{e.trunc.X}
    \normsup{ \exp(L_{\tilde X} f) - \exp(L_{\tilde X^{(L)}}
      f)}{(1-d_1)\delta,(1-d_2)\rho} \prec
    \frac{E}{d_2}\normsup{X}{\delta,1}\normsup{f}{\delta,\rho}\rho^{L}\ .
\end{equation}


\section{Outline of Moser's constructive scheme}
\label{s:MoserProof}

The scheme of the proof here reported is taken from
\cite{Lee12}. Since we are mostly interested in translating the idea
of the proof in a applicable scheme, we omit most of the geometric
details, while focusing on the objects one has to construct (in a
given set of coordinates) and on their manipulation.

\begin{theorem}
Let $(M,\omega)$ a symplectic manifold and $\omega$ be a
non-degenerate closed 2-form in a neighborhood $U$ of $P\in M$. Then
there exists $V$, a neighborhood of $P$, and a set of coordinates
$(y_1,...,y_n,x_1,...,x_n)$ defined in $V$ such that $\omega$ is
represented in the standard form
$$\omega=\sum_{i=1}^ndx_i\wedge dy_i\ .$$
\end{theorem}

\begin{proof}

Let $\omega_0$ be a representation of the $\omega$ in a local chart
$(U_0,\varphi_0)$ around the arbitrary point $P\in M$, hence
$\omega=\varphi_0^*\omega_0$; let $\graff{q_i}_{i=1,\ldots,2n}$ the
coordinates related to this chart, hence
\begin{displaymath}
\omega_0=\sum_{i<j}\omega_{0,ij}(q) dq_i \wedge dq_j\ .
\end{displaymath}
We recall that the coefficients $\omega_{0,ij}$ uniquely define the
antisymmetric matrix $\Omega$ which represents the action of
$\omega_0$ on tangent vectors $X,Y$, namely
\begin{equation*}
\omega_0(X,Y) = X^\top \Omega(q) Y\qquad\qquad
\Omega_{i<j}(q)=\omega_{0,ij}(q)\ .
\end{equation*}
We have to show that there exists a local chart $(U_1,\varphi_1)$
centered in $P$, and a second set of coordinates
$\graff{x_i,y_i}_{i=1,\ldots,n}$ such that
$\varphi_1^*(\omega_1)=\omega$ where $\omega_1$ is the standard
symplectic form on $\RR^{2n}$
\begin{displaymath}
\omega_1=\sum_{i=1}^ndx_i\wedge dy_i = \sum_{i<j}\omega_{1,ij}(q) dz_i \wedge dz_j\ ;
\end{displaymath}
here $z_i=x_i$ for $i=1,\ldots,n$ or $z_i=y_i$ for $i=n+1,\ldots,2n$
and $\omega_{1,ij} = J_{i<j}$.  This is equivalent to saying that there
exists a local change of coordinates $(x,y)=\varphi(q)$ such that
\begin{displaymath}
\varphi^*\omega_1(q)=\omega_1(\varphi(q)) = \omega_0(q)\ .
\end{displaymath}
Hence, we can work directly in a neighborhood $U\subset \R^{2n}$ of
the origin $O=\varphi_0(P)=\varphi_1(P)$. From Darboux's theorem on
vector spaces (see \cite{AbrM08}) there exists a linear and symplectic
change of coordinates (simplectomorphism) such that
$\omega_1|_{O}=\omega_0|_{O}$; this means that $\Omega(0) = J$ where
$J$ represents the standard symplectic form $\omega_1$. We consider
$\omega_0$ written in the ``unknown'' coordinates $\graff{z_i}_i$.

Let $\eta=\omega_1-\omega_0$ be the deformation of $\omega_0$ with
respect to $\omega_1$ (indeed they coincide ath the origin); $\eta$ is
closed and exact \footnote{As we are working on $U$ which is convex
  therefore in particular star-shaped, then Poincar\'e's Lemma holds true,
  i.e. $\eta$ is exact.}. Let $\alpha$ be the 1-form such that
$\eta=-d\alpha$, uniquely defined modulo adding $df$, with
$f:U\to\mathbb{R}$. We can always assume\footnote{It is enough to
  redefine $\alpha=\alpha-\alpha_O$} $\alpha_{O}=0$ and write in
coordinates
\begin{equation*}
\alpha(x)=\sum_{i=1}^{2n} a_i(x) dx_i\qquad\qquad a_i(0)=0\ .
\end{equation*}
We introduce time dependent 2-form on $U$
$$\omega_t(x)=\omega_0+t\eta=t\omega_1+(1-t)\omega_0\qquad \ \forall \ t \in
\RR$$
represented by the matrix
\begin{equation*}
\Omega_t(x) = t J +(1-t)\Omega(x)\qquad\qquad \Omega_t(0)=J\quad\forall t\in\RR\ ,
\end{equation*}
which is invertible, being $\omega_t$ always closed and non-degenerate.

The idea of the proof is to construct a time-dependent vector field
$V_t(q)$ on $U$ whose time-one-flow, $\Psi^t(q)\Big|_{t=1}$, defines
the change of coordinates $(x,y)=\varphi(q)$ which puts $\omega_0$
(and hence $\omega$) into its standard form $\omega_1$, namely
\begin{equation*}
\omega_1(\varphi(q)) = \sum_{i<j}\omega_{1,ij} d\varphi_i(q) \wedge
d\varphi_j(q) = \omega_0(q)\ .
\end{equation*}
This is done by showing that, with a suitable definition of $V_t$, one has
\begin{equation*}
\frac{d}{dt}\omega_t(\Psi^t(q)) = \frac{d}{dt}(\Psi^t)^*\omega_t=0
\end{equation*}
so that $(\Psi^t)^*\omega_t$ is constant $\forall t$, and then
$\omega_1(\Psi^1(q))=\omega_0(\Psi^0(q))=\omega_0(q)$, since $\Psi^0$
is the identity transformation of coordinates $z=(x,y)=q$. The vector
field $V_t(q)$ is obtained as a not unique\footnote{Because $\alpha$,
  and hence the vector $a$, in defined modulo a differential $df$, as
  already noticed.}  solution to the so-called ``Moser equation''
\begin{equation*}
\omega_t(V_t,Y)=\alpha(Y)\qquad\qquad\forall Y
\end{equation*}
or more explicitly
\begin{equation*}
\Omega_t(q) V_t(q) = a(q)\qquad\Rightarrow\qquad V_t(q)=\Omega_t^{-1}(q) a(q)\ .
\end{equation*}
The dynamical system $\dot q=V_t(q)$ defines the flow $\Psi^t(q)$
\begin{equation*}
\frac{d}{dt} \Psi^t(q) = V_t (\Psi^t(q))\ .
\end{equation*}
With such a definition of $V_t$ it is possible to explicitly show
\begin{equation*}
\frac{d}{dt}\omega_t(\Psi^t(q)) =
\frac{d}{dt}\sum_{i<j}\omega_{t,ij}(\Psi^t(q)) d\Psi_i^t(q) \wedge
d\Psi_j(q)\ ,
\end{equation*}
by performing the various derivatives and obtaining in coordinates
that that
\begin{equation*}
\frac{d}{dt}\omega_t(\Psi^t(q)) = (\Psi^t)^*\quadr{\eta(x) + d\alpha(x)}=0\ .
\end{equation*}
\end{proof}


\noindent
{\bf Acknowledgments} T.P. and S.P. thank Vassilis Koukouloyannis for
his hospitality in Samos in September 2022, which inspired useful and
intensive discussions about the dynamics of the Ablowitz-Ladik model.
T.P. has been supported by the MIUR-PRIN 20178CJA2B ``New Frontiers of
Celestial Mechanics: Theory and Applications''.



\begin{thebibliography}{00}



\bibitem{AbrM08}
Abraham, Ralph and Marsden, Jerrold E,
\newblock Foundations of mechanics.
\newblock{\em American Mathematical Soc.}, n. 368, 2008.



\bibitem{CHK19}
Christodoulidi H, Hone ANW and Kouloukas TE
\newblock A new class of integrable Lotka-Volterra systems
\newblock {\em Journal of Computational Dynamics}, 6 (2), 223--237, 2019.

\bibitem{DanSPP22}
V.~Danesi, M.~Sansottera, S.~Paleari and T.~Penati
\newblock  Continuation of spatially localized periodic solutions in discrete NLS lattices via normal forms
\newblock {\em Communications in Nonlinear Science and Numerical Simulations} 108 (4), 2022.

\bibitem{Dar1882}
G.~Darboux,
\newblock Sur le probl\`eme de Pfaff
\newblock {\em Bulletin des sciences math\'ematiques et astronomiques 2 e s\'erie}, tome 6, n. 1, 1882.

\bibitem{Dep69}
A. Deprit,
\newblock Canonical transformations depending on a small parameter, 
\newblock {\em Cel.Mech.} 1, 1969.


\bibitem{GengVY08}
J.~Geng, J.~Viveros and Y.~Yi
\newblock Quasi-periodic breathers in Hamiltonian networks of long-range coupling
\newblock {\em Physica D}, 237, 2008.

\bibitem{Gio22} A. Giorgilli,
\newblock Notes on Hamiltonian Dynamical Systems
\newblock {\em London Mathematical Society Student
    Texts}, Cambridge University Press, April 2022.

\bibitem{GioLS15}
    A. Giorgilli, U. Locatelli and M. Sansottera
    \newblock Improved convergence estimates for the Schr\"oder-Siegel problem
    \newblock {\em AMPA}, 194,    2015.

\bibitem{GioZ92}
A. Giorgilli and E. Zehnder,
\newblock Exponential stability for time dependent potentials. 
\newblock {\em ZAMP} 43, 1992

\bibitem{Grob73}
W. Gr\"obner,
\newblock Serie di Lie e loro applicazioni (italian translation)
\newblock Ed. Cremonese, Roma, 1973.

\bibitem{HeiLW02}
  E. Hairer, C. Lubich and G. Wanner,
  \newblock Geometric Numerical Integration
  \newblock {\em Springer Series in Computational Mathematics}, 31, 2006.

\bibitem{HenKC22}
  D. Hennig and N. I. Karachalios and J. Cuevas-Maraver,
  \newblock The closeness of the Ablowitz-Ladik lattice to the Discrete Nonlinear Schr\"odinger equation
  \newblock {\em Journal of Differential Equations}, Vol. 316, 2022. 

\bibitem{HenKC22b}
  Dirk Hennig and Nikos I. Karachalios and Jesus Cuevas-Maraver,
  \newblock The closeness of localized structures between the Ablowitz-Ladik lattice and discrete nonlinear Schr\"odinger equations:
  Generalized AL and DNLS systems
  \newblock {\em Journal of Math. Physics}, Vol. 63, 4, 2022. 
  
\bibitem{Hor66}
G. Hori,
\newblock Theory of general perturbations with unspecified canonical variables,
\newblock {\em Publ. Astron. Soc. Japan}, 18, 1966.

\bibitem{JohA97}
M.~Johanson and S.~Aubry
\newblock Existence and stability of quasiperiodic breathers in the discrete nonlinear
Schr\"odinger equation
\newblock {\em Nonlinearity}, 10,1997.

\bibitem{JungMW17} A. Junginger, J. Main and G. Wunner,
  \newblock Construction of Darboux coordinates and
  Poincar\'e-Birkhoff normal forms in noncanonical Hamiltonian systems
  \newblock {\em Physica D: Nonlinear Phenomena}, Vol. 348, 2017.


\bibitem{Lee12}
M.J. Lee,
\newblock Introduction to Smooth Manifolds.
\newblock {\em Graduate Texts in Mathematics}, Vol 218, 2nd ed., 2012,   

\bibitem{Mit_etal_23}
  Mithun, Thudiyangal and Maluckov, Aleksandra and Man\ifmmode \check{c}\else \v{c}\fi{}i\ifmmode \acute{c}\else
  \'{c}\fi{}, Ana and Khare, Avinash and Kevrekidis, Panayotis G,
  \newblock How close are integrable and nonintegrable models: A parametric case study based on the Salerno model
  \newblock {\em Phys. Rev. E}, Vol. 107, 2, 2023.

\bibitem{Mos65}
J. Moser,
\newblock On the Volume Elements on a Manifold
\newblock {\em Transactions of the American Mathematical Society},
Vol. 120, No. 2, 1965.

\bibitem{PalP14}
S. Paleari and T. Penati,
\newblock An extensive resonant normal form for an arbitrary large
  {K}lein-{G}ordon model.
\newblock {\em Annali Matematica Pura ed Applicata}, 2014.

\bibitem{PalP19} 
S. Paleari and T. Penati,
\newblock {Hamiltonian Lattice Dynamics}.
\newblock {\em Editorial for the Special Issue ``Hamiltonian Lattice Dynamics'', Mathematics in Engineering}, 1
(4), 2019.

\bibitem{PelPP16} 
  D.E. Pelinovsky, T. Penati and S. Paleari,
  \newblock {Approximation of small-amplitude weakly coupled oscillators with discrete nonlinear Schroedinger equations}
  \newblock {\em Reviews in Mathematical Physics}, 28, n 7, 2016.

\bibitem{SS94}
Sanz-Serna JM,
\newblock An unconventional symplectic integrator of W.Kahan
\newblock {\em Applied Numerical Mathematics}, 16, 245--250, 1994.

\bibitem{TangLCS07}
   Yifa Tang, Jianwen Cao, Xiangtao Liu and Yuanchang Sun,
  \newblock Symplectic methods for the Ablowitz-Ladik discrete nonlinear Schr\"odinger equation
  \newblock  {\em J. Phys. A: Math. Theor.}, n. 40, 2007.

\bibitem{You98}
J. You
\newblock Perturbations of Lower Dimensional Tori for Hamiltonian Systems
\newblock {\em Journal of differential equations} 152, 1999.

\bibitem{Yuan02}
X. Yuan
\newblock Construction of Quasi-Periodic Breathers via KAM Technique
\newblock {\em Commun. Math. Phys.}, 226, 2002.

\end{thebibliography}


\end{document}